\documentclass{jpsj2}

\title
{
Evolution in the split-peak structure across the Peak Effect region in single crystals of $2H$-NbSe$_2$
}

\author
{ 
A. D. {\sc Thakur}$^{1,}$\footnote{Email: ajay@tifr.res.in}, T. V. Chandrasekhar {\sc Rao}$^{2}$, S. {\sc Uji}$^{3,}$\footnote{Email: uji.shinya@nims.go.jp}, T. {\sc Terashima}$^{3}$, M. J. {\sc Higgins}$^{4}$ \\ S. {\sc Ramakrishnan}$^{1}$ and A. K. {\sc Grover}$^{1}$
}

\inst
{
$^1$Department of Condensed Matter Physics and Materials Science\\
Tata Institute of Fundamental Research, Mumbai 400005, India\\
$^2$Technical Physics and Prototype Engineering Division\\
Bhabha Atomic Research Center, Mumbai 400085, India\\
$^3$National Institute for Materials Science, Tsukuba, Ibaraki 305-0003, Japan\\
$^4$Department of Physics and Astronomy, Rutgers University, Piscataway, New Jersey 08855, USA
}

\recdate
{
\today
}

\abst
{
We have explored the presence of a two-peak feature spanning the peak effect (PE) region in the ac susceptibility data and the magnetization hysteresis measurements over a wide field-temperature regime in few weakly pinned single crystals of $2H$-NbSe$_2$, which display reentrant characteristic in the PE curve near $T_c$(0). We believe that the two-peak feature evolves into distinct second magnetization peak anomaly well separated from the PE with gradual enhancement in the quenched random pinning.
}

\kword
{
Peak effect, second magnetization peak, split-peak structure in peak effect, vortex phase transformations, $2H$-NbSe$_2$
}

\begin{document}
\maketitle

\section{Introduction}

The exploration of the magnetic phase (H, T) diagrams in weakly pinned type II superconductors has been an area of intense investigations since the advent of high $T_c$ era \cite{rmp, mikitikbrandt1, kier0, giamarchibhattacharya, andrei, palt1, jpsjsuppl, menonphysica, menonprb, troyanovsky, adt1}. One of the important sub-topics in these studies has been the phenomenon of peak effect (PE) in critical current density ($J_c$), which corresponds to an anomalous increase in the $J_c(H, T)$ close to the upper critical field ($H_{c2}(T)$) \cite{pippard}. The classical PE is construed to be an imprint of (spatial) order-disorder transition in the flux line lattice (FLL) \cite{ling, marchevsky}, and it is widely believed to be caused by the faster collapse of the elasticity of the FLL as compared to the pinning strength of vortex to the underlying quenched random disorder in the atomic lattice as $H \rightarrow H_{c2}(T)$ \cite{pippard}. In addition to this classical PE phenomenon, there have been numerous other observations related to a disorder (i.e., pinning) induced order-disorder transition deep within the mixed state (i.e., far away from the $H_{c2}(T)$ line), which also ramify as an anomalous increase in $J_c(H)$ and appear as the widely reported second magnetization peak (SMP) anomaly in isothermal M-H hysteresis scans \cite{khaykovich}.

One may surmise that, broadly there are three different situations reported in the literature: (i) the PE phenomenon lying close to the $H_{c2}(T)$ line \cite{ssb00}, (ii) the SMP lying deep within the mixed state \cite{khaykovich}, and, (iii) the SMP anomaly and the PE phenomenon in the same isothermal scan, with either the former well separated from the latter \cite{sarkar} or with both lying in juxtaposition leading to a broad anomaly \cite{adt1}. Structural studies related to the PE phenomenon in single crystal samples of the elemental $Nb$ ($T_c(0) \sim 9~K$) via the small angle neutron scattering (SANS) experiments have attracted considerable attention in recent years \cite{ling, forgan, park}. The measurements by Ling {\it et al} \cite{ling} reveal the occurrence of disordering of the FLL across the PE regime in a single crystal sample of $Nb$, which has superheating/supercooling characteristics akin to the first order thermal melting of the FLL. These observations do not appear to stand corroborated in a study by Forgan {\it et al} \cite{forgan}, who found the absence of the PE; instead for a vortex state prepared at $H = 2~kOe$ in an ultrapure single crystal of $Nb$, the validity of the Abrikosov's picture of the ordered FLL was seen to hold upto $99.9 \%$ of the $T_c(H)$ value.

The single crystals of $Nb$ utilized by Ling {\it et al} \cite{ling} and Forgan {\it et al} \cite{forgan} are considered to have different levels of quenched random disorder, which not only influence the intrinsic $H_{c2}(0)$ values in these samples, but, also determine the limiting field value below which the PE phenomenon does not get observed, instead the sharp Bragg spots survive upto the $T_c(H)$ value. Park {\it et al} \cite{park} reported that the said limiting field in the crystal of Ling {\it et al} \cite{ling} is about 0.8~kOe, less than half the value in which Forgan {\it et al} \cite{forgan} had reported their results in a more pure crystal of $Nb$. Mikitik and Brandt \cite{mbprl} have sought to rationalize the different limits for the said field value, in the data of Ling {\it et al} \cite{ling} and Forgan {\it et al} \cite{forgan}, by suitably parametrizing the relative importance of pinning and thermal fluctuations and calculating the pinning induced order-disorder line to be distinct from the thermal melting line. The theoretical framework of Mikitik and Brandt \cite{mbprl} permits the intersection of the two disordering lines for a specific choice of the parameter(s), such that over a part of the (H, T) space, one may witness pinning induced order-disorder line and over the remaining (H, T) space, the Abrikosov's ordered FLL state survives upto the thermal melting line, which in the case of $Nb$ would be within few mK of the $T_c(H)$ line. The dimensionless Ginzburg number, $Gi$, defined as $Gi = \frac{1}{2} \left[ \frac{T_c}{H_c^2 \epsilon \xi_0^3}\right] ^2$ (where the symbols have their usual meanings) \cite{rmp, kier0}, determines the width of the temperature interval over which the thermal fluctuations register their dominance. The $Gi \sim 10^{-10}$ for $Nb$, which makes the FLL melting line to lie within $\sim 10^{-3}~K$ of the $T_c(H)$ line. This in turn renders the detailed studies pertaining to the crossover between the two transition lines in $Nb$ samples a very difficult proposition.

Fortunately, another low $T_c$ superconductor, $2H$-NbSe$_2$ ($T_c(0) \sim 7.2~K$), whose samples have been widely studied in the context of vortex phase diagram studies \cite{giamarchibhattacharya, andrei,  menonphysica, troyanovsky, adt1, marchevsky, kokubo, adt2, andreilatest, hbphysica}, has $Gi \sim 10^{-6}$ and $\frac{j_c(0)}{j_0(0)} \sim 10^{-4}$, where $j_c(0)$ and $j_0(0)$ are the critical and depairing current densities in the limits, $T \rightarrow 0$ and $H \rightarrow 0$ \cite{kier0}. The $Gi$ number for $2H$-NbSe$_2$ is intermediate \cite{hbphysica} to those in elemental $Nb$ and the high $T_c$ cuprates, and the ratio of the current density values in $2H$-NbSe$_2$ are such that their single crystals are typically clean and a sharp PE close to the $T_c(H)$ line can be easily witnessed. In addition, single crystals of $2H$-NbSe$_2$ with varying degree of quenched random pinning can be conveniently prepared. A study of the PE phenomenon in the crystals of $2H$-NbSe$_2$ with slowly varying degree of quenched random disorder, therefore, offers a potential of exploring the interesting crossover between the above stated pinning and thermally induced transition lines. In this context, it is fruitful to state here an inference drawn by Kokubo {\it et al} \cite{kokubo} from their recent experiment on the driven FLL in clean crystals of $2H$-NbSe$_2$. They surmise that the disorder induced transition is well separated from the collapse of the shear rigidity of the FLL, as ascertained from the crossover between coherent and incoherent flow for the driven vortex matter. The velocity dependence of the dynamical melting signatures reveal the connection between the disorder driven and the thermally driven transitions, such that  for a given disorder, the separation between the two transitions surfaces up conveniently in an optimally driven vortex state. If we now look back at the results of a study of the PE phenomenon in three crystals (designated as $A$, $B$ and $C$, with progressively higher amount of quenched random disorder) of $2H$-NbSe$_2$ by Banerjee {\it et al} \cite{ssb00}, we can refocus on the two possible modulations, in close proximity to each other in $J_c(T)$ at some field values (viz., $H=15~kOe$) in the sample $B$. The cleanest sample $A$ exhibits a very sharp PE in the $\chi^{ ~\prime}_{ac}(T)$ response, with no visible structure at all fields ranging from 20~Oe to 20~kOe. The $\chi^{ ~\prime}_{ac}(T)$ response in the sample $C$ on the other extreme, however, exhibits a SMP anomaly well separated from the PE, as revealed by studies performed \cite{adt1, adt2} subsequent to the report by Banerjee {\it et al} \cite{ssb00}. At some field values, the $\chi^{~\prime}_{ac}(T)$ response in the PE region in the sample $C$ displayed a two step fracturing response \cite{ssb04}. The correspondence between fracturing behavior in sample $C$ and partially resolved modulation evident in sample $B$ is not clearly apparent in the results of Banerjee {\it et al} \cite{ssb00}.

In an attempt to overcome this gap, we now present here the results of the PE studies in some more detail in a number of crystals of $2H$-NbSe$_2$ with inherent quenched random pinning lying between those of the sample $B$ and the sample $C$ studied earlier \cite{ssb00}. These crystals exhibit reentrance in $t_p(H)$ ($= T_p(H)/T_c(0)$) data at low fields, a feature seen first in the sample $B$ by Ghosh {\it et al} \cite{kghosh}. The observations in the newer crystals, unravel the existence of a well resolved two-peak feature across the PE in the intermediate field region ($1~kOe < H < 36~kOe$; $H~\|~c$). At higher fields, the field induced (i.e., higher effective disorder) broadening appears to coalesce the two-peak feature into an unresolved broad peak centered around 37~kOe (at 50 mK; $H~\|~c$). The investigations of the PE for $H~\|~ab$ reveal a well resolved two-peak feature even at fields in the range of 105 to 120~kOe at 50~mK. We believe that a two-peak feature so observed could be an essential attribute of the weakly pinned vortex matter, as it evolves eventually into well researched notions of second magnetization peak (SMP) anomaly and PE, with progressive increase in the quenched random pinning.

\section{Experimental}

We present data on ac susceptibility measurements in two single crystals of $2H$-NbSe$_2$, named here as sample $B2$ and sample $B3$, both of them with $T_c(0) \sim 7.2 K$. We also present few ac susceptibility results in the crystal $B$ used by Banerjee {\it et al} \cite{ssb00}, and in the crystal $C^{\prime}$ (which is same as the sample called $Z^{\prime}$, studied by Thakur {\it et al} \cite{adt1, adt2}, whose behavior is analogous to that in sample $C$ of Banerjee {\it et al} \cite{ssb00}). In addition, some of the data from isothermal M-H measurements is also being presented in sample $B2$. The residual resistance ratio ($\frac{R_{300 K}}{R_{8 K}}$) in the samples $A$, $B$ and $C$ were 20, 16 and 9, respectively \cite{ssb00}. Banerjee {\it et al} \cite{ssb00} had estimated the value of the ratio $\frac{R_c}{a_0}$ at 10~kOe ($\|c$) in sample $A$ to be 400, which reduces to about 20 for sample $C$, here $R_c$ stands for the radial correlation length of FLL and $a_0$ is the lattice constant ($\approx 500~\AA$ at 10~kOe). We find the residual resistance ratio ($\frac{R_{300 K}}{R_{8 K}}$) in samples $B3$ and $C^{\prime}$ to be 13.9 and 9.8 respectively.

Low and intermediate field ac susceptibility measurements were done utilizing (i) an ac option facility on a commercial Quantum Design Inc., U.S.A., SQUID magnetometer (Model MPMS7) and (ii) the usual astatic coil arrangement in a home built \cite{ramky} ac susceptibility setup, in the frequency interval 1~Hz to 1111~Hz, and for ac field ($h_{ac}$) amplitudes lying in the range of 0.5~Oe to 3.0~Oe (r.m.s.). High field ac susceptibility measurements were done inside a top loading dilution refrigerator, fitted with a 16~Tesla superconducting magnet. Isothermal M-H loops have been recorded on a 12~Tesla Vibrating Sample Magnetometer (VSM) (Oxford Instruments, U.K.) at a field ramp rate of 4~kOe/min.

\section{Results}

\subsection{In phase ac susceptibility measurements}

Imprints of the variations in the critical current density $J_c$ can be captured via the ac susceptibility measurements: $\chi^{~\prime}_{ac} \sim -\beta J_c(H, T)/h_{ac}$ for $h_{ac} > h^{\star}$ and $\chi^{~\prime}_{ac} \sim -1 + \alpha h_{ac} / J_c(H, T)$ for $h_{ac} < h^{\star}$, where $\alpha$ and $\beta$ are geometry and size dependent factors and $h^{\star}$ is the parametric field at which the induced screening currents flow through the entire sample \cite{lingbud}.

\subsection{Isofield $\chi^{~\prime}_{ac}(T)$ data for $H~\|~c$ in the crystals $B$ and $C^{\prime}$ of $2H$-NbSe$_2$}

Fig. 1 shows a comparison of the isofield ac susceptibility response, $\chi_{ac}^{\prime}(T)$ for the samples $B$ and $C^{\prime}$ (note that the latter is stronger pinned as compared to the former). In sample $B$ (Fig. 1 (a)), the PE is a single sharp  dip in $\chi^{~\prime}_{ac}(T)$ centered at 6.72 K for $H = 2~kOe$ (FLL constant, $a_0 \sim 1100 \AA$), while for sample $C^{\prime}$ (Fig. 1 (b)), there is a characteristic two step disordering evident across the PE region in $H = 3~kOe$ ($a_0 \sim 900 \AA$). In addition, one observes another anomalous variation prior to the PE centered at 4.73~K in the case of sample $C^{\prime}$, which we label as the fingerprint of SMP anomaly \cite{adt2}. An  inset panel in Fig. 1 (a) recalls \cite{ssb00} the data in sample $B$ at $H = 15~kOe$ ($a_0 \sim 400 \AA$), the partially resolved structure in $\chi^{~\prime}_{ac}(T)$ response across the PE can be imagined to be a juxtaposition of two anomalous variations in $J_c(T)$. For a given amount of quenched random disorder, the effective disorder is expected to enhance \cite{menondasgupta}, as $a_0$ decreases in response to increase in $H$ ($a_0 \propto H^{-0.5}$). Banerjee {\it et al} \cite{ssb00} had attributed the broadening of the PE at higher $H$ to an enhancement in effective pinning.  

\subsection{Isofield $\chi^{~\prime}_{ac}(T)$ data at low fields ($H~\|~c$) in crystal $B2$ of $2H$-NbSe$_2$}

Panel (a) in Fig. 2 shows low field $\chi^{~\prime}_{ac}(T)$ data ($H~\|~c$) recorded using a SQUID magnetometer in the field range, $50~Oe < H < 1~kOe$ ($7000 \AA > a_0 > 1500 \AA$), for the crystal $B2$, which is the main focus of the present report. On examining the evolution in the $\chi^{~\prime}_{ac}(T)$ curves recorded for fields ranging from 1000~Oe to 150~Oe, it can be noted that the dip feature in the susceptibility response becomes more prominent as $H$ decreases from 1000~Oe to 300~Oe, while the $T_p(H)$ values increase. However, as the applied field is reduced below 250~Oe, the dip feature progressively broadens  and the $T_p(H)$ values of broad anomaly appear to decrease with further decrease in field. Fig. 2 (b) shows a plot of the ratio $T_p(H)/T_c(0)$ ($\equiv$~reduced temperature, $t_p$) {\it vs} $H$ in crystal $B2$ of $2H$-NbSe$_2$. The onset of the non-monotonicity in $t_p(H)$ behavior below 200~Oe identifies the reentrant characteristic in the PE curve, analogous to that in the crystal $B$ \cite{kghosh}. Another facet which elucidates the similarity between the present crystal $B2$ and the earlier crystal $B$ is the evolution of PE in $\chi^{~\prime}_{ac}(T)$ response between 500~Oe and 1000~Oe. Such a tendency, however, shows a reversal as the field starts to increase beyond 4~kOe $-$ this would become evident from the  $\chi^{~\prime}_{ac}(H)$ plots shown in Fig. 3.

\subsection{Isothermal $\chi^{~\prime}_{ac}(H)$ data in crystal $B2$}

In Fig. 3 (a), the $\chi^{~\prime}_{ac}(H)$ response at 6.95~K depicts PE as a broad unresolved dip centered around $H_p \sim 800~Oe$. Such a behavior appears to be consistent with the $\chi^{~\prime}_{ac}(T)$ plot in $H = 1000~Oe$ in Fig. 2 (a). However, as the temperatures get lowered, the $\chi^{~\prime}_{ac}(H)$ responses at 6.9~K, 6.85~K and 6.8~K in Fig. 3 (a) start to display a resolved split-peak feature. This trend can be seen to continue in Fig. 3 (b) and Fig. 3 (c) as the temperatures decrease from 6.5~K to 3.5~K, and the field region of PE progressively shifts to higher field values. The isothermal curve at 6.5~K shows the occurrence of PE feature in the field interval of 3.4 to 4~kOe, where the normalized $\chi^{~\prime}_{ac}(H)$ values are in the range of just $1$ to $2~\%$ of the perfect shielding response. At lower temperatures ($T < 5.5~K$), the values of the $\chi^{~\prime}_{ac}(H)$ signals across the PE region appear to once again progressively enhance and the width of this region also increases ({\it cf.} curves in Fig. 3 (c)).

Fig. 4 shows the $\chi^{~\prime}_{ac}(H)$ data in the crystal $B2$ (for $H~\|~c$) between 2~K and 50~mK in $h_{ac}$ of 3~Oe (r.m.s.) using a double coil arrangement installed inside a dilution refrigerator. At 2~K, one can observe the presence of a well resolved split-peak feature centered around 32~kOe. As the temperatures progressively decrease from $2~K$ to $0.55~K$, the width of the PE region continues to enhance. At 50~mK, the PE once again imprints as a broad (barely resolved) dip, a behavior reminiscent of the $\chi^{~\prime}_{ac}(H)$ plot at 6.95~K in Fig. 3 (a). At 50~mK, the onset of the PE commences around $H = 34~kOe$ ($a_0 \approx 270 \AA$), whereas at 6.95~K, it sets in around 500~Oe ($a_0 \approx 2200 \AA$). We shall argue later that the broadening of the PE region witnessed at lower fields near $T_c(0)$ and that at higher fields near $H_{c2}(0)$ is caused by the enhancement in the effective disorder in a given sample at both these ends.

\subsection{Isofield $\chi^{~\prime}_{ac}(T)$ data at higher fields in crystal $B2$}

The isofield plots in $H \leq 500~Oe$ in Fig. 2 (a) showed that PE imprints as a composite single peak with $T_p$ values $> 6.9~K$. To expound the behavior at higher fields, we show in Fig. 5 the representative $\chi^{~\prime}_{ac}(T)$ data in $H = 25~kOe$ and $H = 3.5~kOe$ recorded in $h_{ac}$ of 3.0~Oe (r.m.s.) using the SQUID magnetometer. In both the $\chi^{~\prime}_{ac}(T)$ curves in Fig. 5, PE shows up as a split two-peak structure. In consonance with the trend evident in $\chi^{~\prime}_{ac}(H)$ data (see Fig. 3(b) and Fig. 3(c)), the split two-peak feature is well resolved for $H = 25~kOe$, where the onset of PE ($T_p^{on}$) occurs at about 3.1~K. The same splitting is less conspicuous for $H = 3.5~kOe$, where $T_p^{on}$ is around 6.5~K in the inset panel of Fig. 5.

A comparison of the $\chi^{~\prime}_{ac}(T)$ plots in crystal $B$ at 2~kOe and 15~kOe in Fig. 1 (a) with those in crystal $B2$ at 1~kOe, 3.5~kOe and 25~kOe in Fig. 3(a) and Fig. 5 leads to the premise that progressive increase in effective pinning initially results in broadening of the PE, and it eventually resolves the structure across it.

\subsection{Isofield $\chi^{~\prime}_{ac}(T)$ data in crystal $B3$ of $2H$-NbSe$_2$}

To strengthen the above premise, we show in Fig. 6, the isothermal $\chi^{~\prime}_{ac}(T)$ plots in another crystal ($B3$) of $2H$-NbSe$_2$ at $H = 1.5~kOe$, $3~kOe$ and $15~kOe$, respectively. Note first that the split-peak feature across the PE at 15~kOe is clearly resolved into two separate peaks in crystal $B3$ in Fig. 6. This may be contrasted with the partially resolved $\chi^{~\prime}_{ac}(T)$ response at 15~kOe in the crystal $B$ (see inset panel in Fig. 1 (a)). Just as the reduction in field from 15~kOe to 2~kOe in the crystal $B$ in Fig. 1 (a) results in coalescing the barely resolvable split peak feature into a single peak, the progressive reduction in the field from 15~kOe to 1.5~kOe in crystal $B3$ manifests as a movement of the precursor peak towards the second peak located at the edge of the (bulk) irreversibility temperature ($T_{irr}^{bulk}$) as shown in the different panels of Fig. 6. The similarity in $\chi^{~\prime}_{ac}(T)$ response at 1.5~kOe in crystal $B3$ with that in crystal $B$ at 15~kOe (cf. Fig. 6 (a) and inset panel of Fig. 1 (a)) is most instructive. At further lower fields ($H < 1~kOe$), the PE in crystal $B3$ manifests as a composite single peak (data not being shown in Fig. 6 for this crystal; however, see plots in Fig. 2(a) for the crystal $B2$).

\subsection{Identification of the onset position of PE and the collapse of shear rigidity}

A split two-peak structure in the $\chi^{~\prime}_{ac}$ response is probably originating from the onset of multiple (at least two of them) transformations in the ordered FLL. It is tempting to identify one of these transformations commencing deeper in the mixed state with the pinning induced disordering and the other (later) with the disordering related to the incipient softening of the elastic moduli of the ordered lattice. Once the shear rigidity of the vortex solid collapses, the pinning characteristic in the bulk of the sample rapidly decreases, presumably due to softening of the normal cores of the vortices on approaching the normal state boundary. The small angle neutron scattering experiments in conjunction with the transport measurements in the weakly pinned crystals of $2H$-NbSe$_2$ have earlier indicated \cite{pautrat} that the surface pinning continues to survive well above the peak position of the PE, i.e., after the pinning in the bulk of the sample vanishes across the PE. It is straight-forward to mark the field/temperature values corresponding to the onset positions ($H_p^{on}$, $T_p^{on}$) of the PE in the isothermal/isofield $\chi^{~\prime}_{ac}$ plots in Fig. 3, Fig. 4 and Fig. 5. It is also fruitful to attempt to identify the field/temperature values marking the vanishing of the shear rigidity of the ordered lattice. A surmise made by Kokubo {\it et al} \cite{kokubo} prompts us to assert that the field/temperature values at which the pinning in the bulk ceases could notionally define such a limit. Taking a cue from an analysis presented by Pautrat {\it et al} \cite{pautrat}, we propose to identify the collapse of (bulk) pinning by an extrapolation procedure as displayed in the Fig. 3 (c) and Fig. 6 (c). The $T_{irr}^{bulk}$/$H_{irr}^{bulk}$ values determined via such an extrapolation probably identify the limits above which the surface pinning (and consequently residual irreversibility in the magnetization hysteresis response) still survives. The arrows at different positions in Fig. 3, Fig. 4 and Fig. 5 mark out the respective onset and end positions of the PE region. An earlier statement implies that the former ($H_p^{on}$ / $T_p^{on}$) may be identified with the start of the pinning induced disordering and the latter ($H_{irr}^{bulk}$/$T_{irr}^{bulk}$) may be (notionally) identified with the vanishing of the shear rigidity. The interval between the two limits is determined by effective pinning. In the limit that the pinning is very nascent, as in the crystal $A$ of $2H$-NbSe$_2$ studied by Banerjee {\it et al} \cite{ssb00}, the onset ($T_p^{on}$), the peak ($T_p$) and the irreversibility ($T_{irr}^{bulk}$) temperatures are so close (within 0.5~$\%$) that the thickness of the PE boundary in the vortex phase diagram sketched by them \cite{ssb00} encompasses the interval, $T_p^{on}$ to $T_{irr}^{bulk}$, in the said sample.

\subsection{Frequency dependence in $\chi^{~\prime}_{ac}(T)$ data and the M-H loops in crystal $B2$}

There could be a concern that the surface/edge effects, which relate to the injection of disordered bundles of vortices from the edges \cite{palt1} and their annealing into the ordered state in the interior may be responsible for the observed structure across the peak effect \cite{xiao00} in $\chi^{ \prime}_{ac}(T) /  \chi^{ \prime}_{ac}(H)$ measurements. To evaluate the validity of such an apprehension, we explored the frequency dependence in $\chi^{ \prime}_{ac}(T)$ response in crystal $B2$ for $H = 12~kOe$ ($\|~c$) in $h_{ac}$ of 1~Oe (r.m.s.) in the frequency range 1~Hz to 1111~Hz, using the ac option facility in the SQUID magnetometer. The results displayed in Fig. 7 (a) show that the two-peak structure remains well resolved (and unaltered) over the three orders of frequency range investigated. The variation in frequency does not preferentially affect one peak {\it w.r.t.} the other. If the precursor peak had relationship with the surface currents, this would have been affected by the change in frequency from 1~Hz to $\sim 10^{3}$~Hz (as at higher frequencies, the contribution from the edge contamination is preferentially very much reduced \cite{palt1, xiao00}). 

To reaffirm the belief that a two-peak structure across PE relates to the transformation in the state of the vortex matter in the bulk, we recorded isothermal M-H loops at different temperatures. Fig. 7 (b) shows a portion of the M-H loop recorded at 2.5~K using a VSM. The inset panel in Fig. 7 (b) focuses attention onto the PE region, where a split two-peak feature manifests as modulations in the width of hysteresis ($\Delta M(H)$ {\it versus} $H$). Fig. 7 (c) shows a trace of the M-H loop across the PE region at 6.75~K, a modulation in the width of the loop {\it versus} $H$ is evident here as well. The hysteresis width in the M-H loops directly relates to the macroscopic currents set up in the bulk of the sample \cite{fietzandwebb}. We have identified the limit of collapse of irreversible behavior in the bulk of the sample in Fig. 7 (c) in the spirit of marking of the $H_{irr}^{bulk}$/$T_{irr}^{bulk}$ values in $\chi^{~\prime}_{ac}$ data, as described in the previous section.

\subsection{In-phase ac susceptibility data for $H~\|~ab$ in crystal $B2$}

The crystals of the hexagonal $2H$-NbSe$_2$ grown by the vapor transport method are typically platelet shaped, with c-axis normal to the plane of the platelet. While performing magnetization measurements for $H~\|~c$ in such platelets, the demagnetization factor ($N$) is typically large ($N > 0.8$). The effects and complications related to a large $N$ value could get appreciably reduced if the magnetization measurements are performed for $H~\|~ab$. The intrinsic and material superconducting parameters of $2H$-NbSe$_2$ have a large anisotropy, e.g., $H_{c2}(\|~ab) / H_{c2}(\|~c) \sim 3$. Keeping these in view, let us now examine in Fig. 8 the $\chi^{ \prime}_{ac}(H)$ data recorded at 50~mK and 2.2~K for $H~\|~ab$ in the crystal $B2$ of $2H$-NbSe$_2$. A well resolved two-peak structure is evident at both these temperatures, the PE region appears broader at 50~mK as compared to that at 2.2~K. This attests to the notion of enhancement in effective pinning at higher fields \cite{ssbthesis} even for $H~\|~ab$. It is satisfying to find that the splitting in the two-peak structure remains well resolved even when applied field is $\sim 110~kOe$ ($a_0 \approx 150 \AA$). The width of PE region at 50~mK for $H~\|~ab$ is about thrice that for $H~\|~c$ and the separation between the two peaks also scales by the same factor. This permits the resolution of the two peaks in the former case as the broadening of the individual peaks does not scale by factor of three.

\subsection{Vortex phase diagram for $H~\|~c$ in sample $B2$}

We show in Fig. 9 a field-temperature (H, T) phase diagram in the crystal $B2$ for $H~\|~c$, where we have collated the field/temperature values of the onset position of the PE, the extrapolated values of the (bulk) irreversibility temperature/field ($T_{irr}^{bulk} / H_{irr}^{bulk}$) and the values of the upper critical field, $H_{c2}(T)$, as ascertained from the onset of the diamagnetic response in the $\chi^{ \prime}_{ac}(H)$ / $\chi^{ \prime}_{ac}(T)$ data. We have also (notionally) indicated the sliver region of reentrant disordered phase at low fields, which lies in between $H \sim 200~Oe$ and $H_{c1}(T)$ (for $H~\|~c$) \cite{ssb00, ssbthesis}. The phase space in between the reentrant disordered phase boundary and the $H_p^{on}$ line could be identified with the well ordered elastic glass \cite{gialed} regime (i.e., the Bragg Glass phase \cite{gialed}). The (H, T) region between $H_p^{on}$ and $H_{irr}^{bulk}$ could be termed as the phase co-existence region, where the ordered weaker pinned regions co-exist with the disordered stronger pinned regions. 

\section{Discussion}

We have presented experimental data obtained from the ac susceptibility and the dc magnetization hysteresis measurements with a view to support an assertion of observation of two distinct happenings \cite{kokubo} across the peak effect in the critical current density over a wide field-temperature regime in weakly pinned single crystals of $2H$-NbSe$_2$, which have specific amount of quenched random disorder, as in their samples $B2$ and $B3$. We have also shown that crystal $B2$ displays the reentrant characteristic in the locus of the peak temperature, $T_p(H)$. Such a behavior had earlier been witnessed in few specific samples of $2H$-NbSe$_2$ \cite{ssb00, kghosh}. The notion of the reentrant characteristic in $T_p(H)$ is rationalized by invoking the onset of the disorder induced broadening of the PE regime, and it is considered to relate to the transition/crossover from small bundle (or individual pinning) regime to the collectively pinned elastic glass regime \cite{ssb02}. The crystals of $2H$-NbSe$_2$, which displayed \cite{ssb00, kghosh} reentrance in $T_p(H)$ at low fields, also showed (see Fig. 1 (a)) significant broadening (and partially resolved structure) \cite{ssb00} in the PE phenomenon at higher fields ($H > 10~kOe$; $a_0 < 500 \AA$). Much more stronger pinned crystals (e.g., sample $C$ or $C^{\prime}$) of $2H$-NbSe$_2$ ($T_c(0) \sim 6~K$) did not display \cite{adt1, ssb00, adt2} PE at low fields ($H < 400~Oe$; $a_0 > 2500 \AA$), however, they did reveal \cite{adt1, adt2} the presence of a precursor peak (named as the SMP in Fig. 1 (b)) prior to the PE at higher fields ($H > 1~kOe$), in addition to the characteristic stepwise disordering feature across the PE. The crystals $B2$ and $B3$, the focus of the present work, presumably have quenched random pinning intermediate to those in the crystal $B$ and the crystal $C$. The reentrant disordered phase \cite{adt1, ssb00, adt2} at low fields in crystal $C$ spans a much larger (H, T) region as compared to that in sample $B$. In crystal $B2$, the reentrant disordered state occupies (H, T) phase space intermediate to those in the crystals $B$ and $C$. This stands substantiated by the fact that Ghosh {\it et al} \cite{kghosh} had observed PE in $\chi^{ \prime}_{ac}(T)$ scans in fields down to $\sim 30~Oe$ in sample $B$, whereas in the crystal $B2$, fingerprint of PE is not evident for $H < 150~Oe$ (cf. Fig. 2). A flat demarcation line (dotted line in Fig. 9), below which the reentrant disordered state prevails, moves progressively downwards (in field) as the pinning reduces from sample $C$ (or $C^{\prime}$) to sample $B$ through sample $B2$ (or $B3$).

It is instructive to compare the vortex phase diagram in Fig. 9 in crystal $B2$ with the vortex phase diagram in crystal $C^{\prime}$  (i.e., sample $Z^{\prime}$ ) of $2H$-NbSe$_2$ (see Fig. 6 in Ref. \cite{adt2}), where the phase co-existence region can be conveniently bifurcated into two parts (I and II), in which the ordered and the disordered pockets, respectively, dominate the bulk electromagnetic response. In the context of the crystal $C^{\prime}$, it has been argued that in the first part (I), the vortex matter is in the polycrystalline form (i.e., multi-domain vortex glass) \cite{moretti, chandran, fasano, menghini, ssbdae}. The above stated flat demarcation line in sample $C^{\prime}$ (or sample $B$) is made up of values of $H_{plateau}(T)$, which are determined from normalized plots of critical current density versus normalized field as described by Banerjee {\it et al} \cite{ssb02}. For $H \leq H_{plateau}(T)$, the vortices are in small bundle collective pinning regime, and as the field values lower further ($H \ll H_{plateau}$), a crossover to the individual pinning regime occurs, where $J_c(H)$ is nearly independent of H ($\geq H_{c1}$). To rephrase, the interval between $H_{c1}$ and $H_{plateau}$ can be imagined to subdivide into two portions, such that in the lower portion, the reentrant disordered phase is completely amorphous and at somewhat higher values of $H$, the polycrystalline form of vortex glass, as imaged \cite{fasano, menghini} in Bitter decoration experiments, can emerge. The potential of multi-domain vortex state to yield PE phenomenon (and the possible structure across it) in $\chi^{ \prime}_{ac}(T)$ scans would not only depend on the (average) size of the domains, but also, on the width of their size distribution and a variety in morphology (shapes) of the domains/domain walls. If the average domain size is small and/or there is a large variation in the shape of the domains, the possibility of observing PE phenomenon in bulk $\chi^{ \prime}_{ac}(T)$ response of the sample would diminish. This could rationalize as to why Fasano {\it et al} \cite{fasano} did not observe PE in some specimen of $2H$-NbSe$_2$ at low fields, which had yielded polycrystalline state in Bitter decoration images. At this juncture, it is fruitful to recall that numerical simulations have revealed that notion of entropic change associated with thermal melting of atomic clusters gets compromised for small sized clusters having a large variety in their shapes \cite{kanhere}. An order-disorder phenomenon associated with PE is considered a precursor to thermal melting of FLL \cite{kwok}.

In sample $C^{\prime}$, the polycrystalline form of vortex matter at higher fields ($H >$ few kOe), where the interaction between the vortices is fully effective (i.e., $a_0 \ll \lambda$, where $\lambda$ is the penetration depth), comprises large  enough domains in which vortices are collectively pinned and elastic behavior prevails. Imposition of external driving forces (e.g., large $h_{ac}$ field) can improve the state of spatial order in the polycrystalline form at a given $H$ by shrinking the domain wall regions \cite{adt2}. The spatial order has also been observed to improve with enhancement in temperature and/or field in between the peak position of the SMP anomaly and the onset position of the PE \cite{capramana}. The dislocations permeating between onset and peak positions of SMP can get (partially) annealed (i.e., squeezed out)  thereafter upto the arrival of the onset of PE. However, as the domain sizes start to shrink on going across the onset position of PE into the second part (II) of the phase co-existence region, the external driving forces (e.g., even large ac field) start aiding the progress towards the complete amorphization of the vortex matter. Large domains can reveal stepwise amorphization between onset and peak positions of the PE for specific combination of field (i.e., intervortex distance $a_0$ within the domain), correlation volume of the domain and the driving force $h_{ac}$, as is evident in the studies of the effect of pinning and drive in sample $C^{\prime}$ of $2H$-NbSe$_2$ \cite{adt2} and the samples of few other low $T_c$ systems (e.g., CeRu$_2$ \cite{adt2}, Yb$_3$Rh$_4$Sn$_{13}$ \cite{ybpramana}, etc.). A large ac driving force has the potential to compromise the sudden change(s) in $\chi^{ \prime}_{ac}(T)$ response (e.g., fracturing notion) between the onset and the peak position of PE. The presence of structure across the PE does not preclude the occurence of an anomalous variation (corresponding to crossing of the SMP line in vortex phase diagram) prior to reaching the PE region, if the driving force is not too large \cite{tulapurkar}. Simultaneous measurement of ac susceptibility and dc resistivity in sample $C$ of $2H$-NbSe$_2$ do indicate that SMP anomaly in $\chi^{ \prime}_{ac}(T)$ response stands (preferentially) progressively suppressed as the transport current increases \cite{tulapurkar}. The precursor SMP peak signals the break up of a Bragg glass (BG) like phase into multi-domain vortex glass (VG) phase. In the sample $B2$, the precursor peak and the peak signaling the collapse of (bulk) pinning lie in juxtaposition to each other above the limiting field-temperature region of reentrant disordered phase. In crystal $B$, the said two peaks overlap to give a narrow composite peak for $H < 5~kOe$, thereafter, the composite peak first broadens and it gradually starts to display some structure on further increase in field. the similarity in the structure of $\chi^{ \prime}_{ac}(T)$ curves across PE region in sample $B$ at 15~kOe and at 1.5~kOe in sample $B3$ (cf. Fig. 1 (a) and Fig. 6 (a)) not only elucidates the role of progressive enhancement in quenched random disorder but also exemplifies the notion of enhancement in effective disorder with increase in field for a given amount of quenched disorder. A well resolved two peak structure across PE region in $\chi^{ \prime}_{ac}$ response that surfaces in samples $B2$ and $B3$ at few kOe survives upto 35~kOe ($H~\|~c$) and down to 1.4~K. At lower temperatures ($T \rightarrow 0$), where PE happens at still higher fields, further enhancement in effective disorder broadens the two peaks into a composite anomaly, where shape resembles the $\chi^{ \prime}_{ac}(H)$ behavior evident at temperatures above the nose region ($T \geq 6.95~K$), i.e., where the flat reentrant phase boundary appears to meet the $H_p^{on}(T)$ line in Fig. 9.

\section{Conclusion}

To conclude, the results presented in Fig. 3 through Fig. 8 amount to propagating the notion that two anomalous variations in $J_c$ are a characteristic attribute of the weakly pinned vortex matter. A distinct identification of a precursor peak and its separation from the quintessential PE peak (with or without additional structure across it) depends on the quantum of effective pinning in a given circumstance (e.g., in presence of a driving force). It had been stated in the introduction that two anomalous variations in the critical current density, named as SMP and PE, stand well documented in the recent literature in several low $T_c$ and high $T_c$ compounds, e.g., Ca$_3$Rh$_4$Sn$_{13}$ \cite{sarkar}, YNi$_2$B$_2$C and LuNi$_2$B$_2$C \cite{djn}, V$_3$Si \cite{kupfer}, YBa$_2$Cu$_3$O$_7$ \cite{sarkar, dpal, nishizaki}, etc. In that context it is also pertinent to recall a recent report \cite{kobayashi} of two-peak structure in the flux-flow resistivity ($\rho(H)$) data in an unconventional superconductor, PrOs$_4$Sb$_{12}$ ($T_c(0) = 1.85~K$ and $H_{c2}(0) = 2.45~T$). Prima facie, the flux flow resistivity data in this crystal (see Fig. 1 and Fig. 2 of Kobayashi {\it et al} \cite{kobayashi}) shows two anomalous variations in $J_c(H)$, akin to the $\chi^{ \prime}_{ac}(H)$ behavior displayed in the panels of our Fig. 3. The underlying physics issues in PrOs$_4$Sb$_{12}$ and $2H$-NbSe$_2$ are however different. In the former, an analysis of angle resolved magnetothermal transport data \cite{matsuda} has revealed a change in symmetry of superconducting order parameter from 4-fold to 2-fold deep inside the mixed state. If we allow for a coupling between the symmetry of the flux line lattice and that of the superconducting gap function, the above could imply a change in FLL symmetry (and its pinning characteristics) across a phase boundary marking transition from 4-fold to 2-fold symmetry, thereby rationalizing the first anomaly in $\rho(H)$ data in PrOs$_4$Sb$_{12}$ compound. The higher field anomaly can then be associated with the usual order-disorder transition, as in $2H$-NbSe$_2$. The state of spatial order could also undergo a change (i.e., deteriorate) across a 4-fold to 2-fold phase boundary. Thus, in the end, we may state that a modulation in the relative fraction of ordered and disordered regions across a large portion of the mixed state appears to emerge as a generic behavior in superconductors showing multiple anomalous variations in its critical current density.

\section{Acknowledgment}

We would like to gratefully acknowledge S. S. Banerjee for access to the data in the crystal $B$. We would also like to thank D. Pal and D. Jaiswal-Nagar for various discussions. One of us (ADT) would like to acknowledge the TIFR Endowment Fund for the Kanwal Rekhi Career Development Fellowship.

\newpage
\begin{figure}
\begin{center}
\includegraphics[width=12cm]{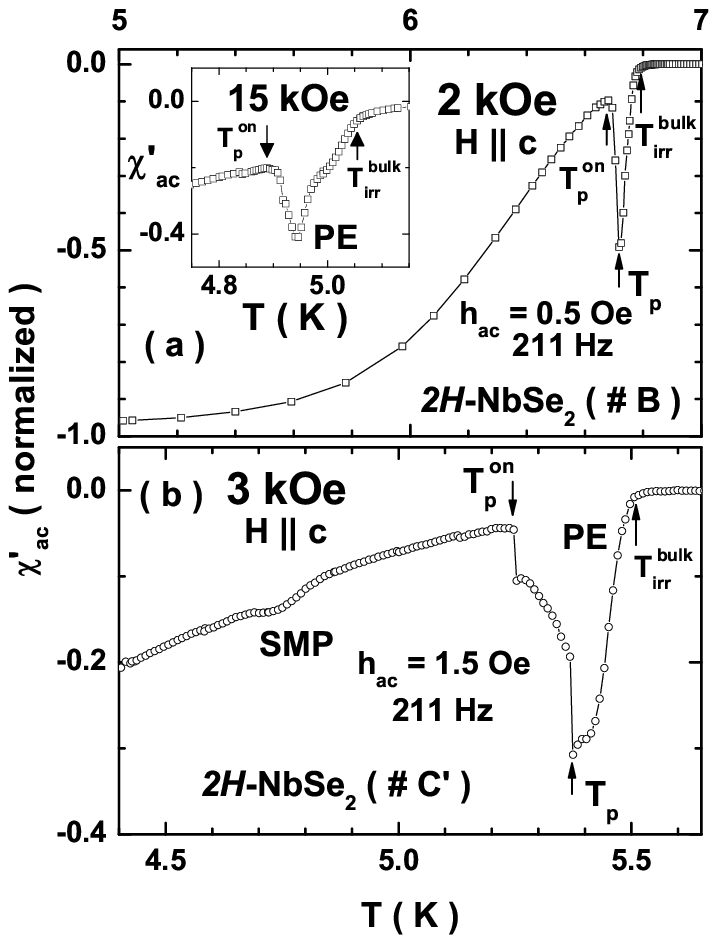}
\end{center}
\caption{In-phase iso-field ac susceptibility ($\chi^{\prime} (T)$) data at 2~kOe ($H~\|~c$) and 3~kOe ($H~\|~c$) in : (a) crystal $B$ ($T_c(0) \approx 7.25 K$) and (b) crystal $C^{\prime}$ ($T_c(0) \approx 6.0 K$) of $2H$-NbSe$_2$.}
\label{fig:1}
\end{figure}
 
\begin{figure}
\begin{center}
\includegraphics[width=12cm]{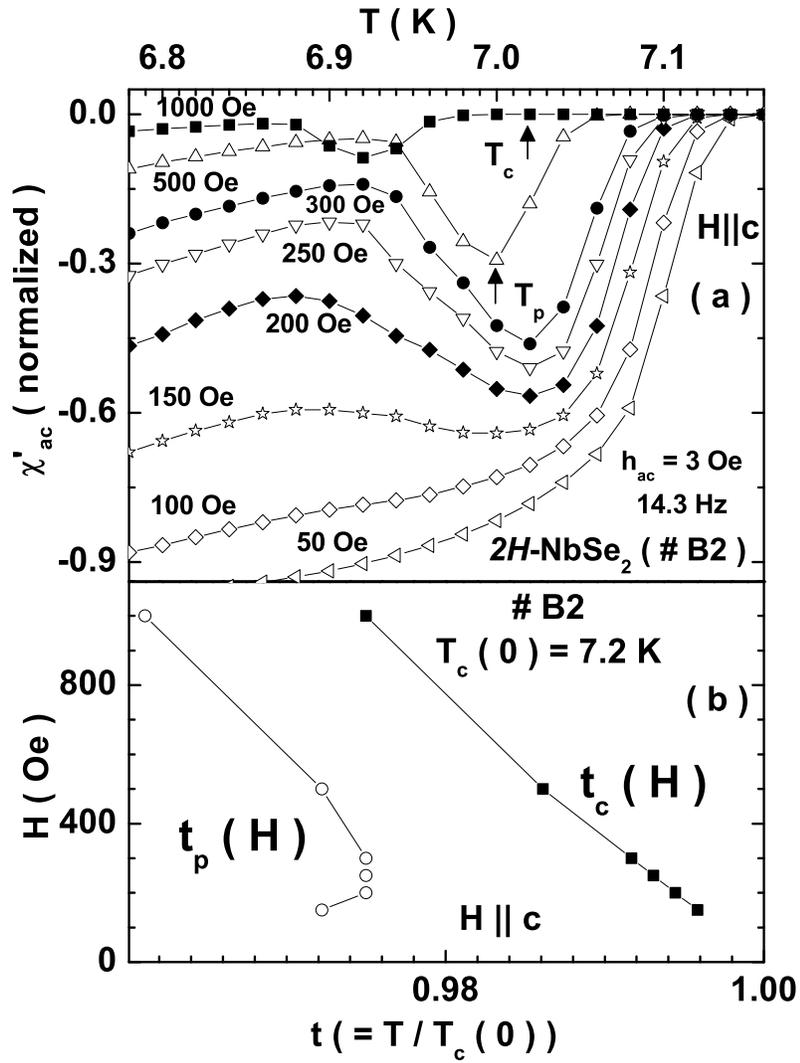}
\end{center}
\caption{Panel (a) depicts the isofield ($H~\|~c$) ac susceptibility response, $\chi_{ac}^{\prime}(T)$, for crystal  $B2$ at low fields ($H~=~50~Oe~to~1000~Oe$). Panel (b) shows the low field portion of a field-temperature (H, t) diagram in the crystal $B2$ constructed from $T_p(H)$ and $T_c(H)$ data in the panel (a).}
\label{fig:2}
\end{figure}

\begin{figure}
\begin{center}
\includegraphics[width=12cm]{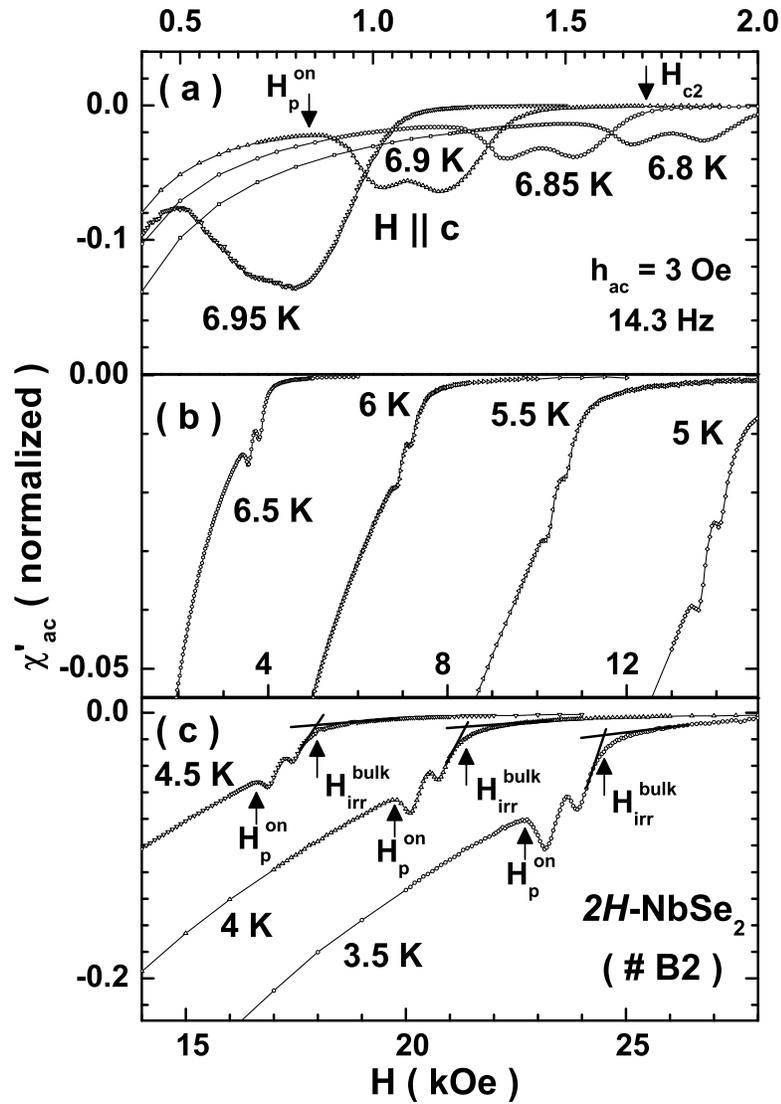}
\end{center}
\caption{Isothermal in-phase ac susceptibility ($\chi_{ac}^{ \prime}(H)$) data ($H~\|~c$), at the temperatures ranging from close to $T_c(0)$ down to 3.5~K, displaying two anomalous variations in $\chi_{ac}^{ \prime}(H)$, which in turn indicates two anomalous variations in $J_c(H)$ in the crystal $B2$ of $2H$-NbSe$_2$.}
\label{fig:3}
\end{figure}

\begin{figure}
\begin{center}
\includegraphics[width=12cm]{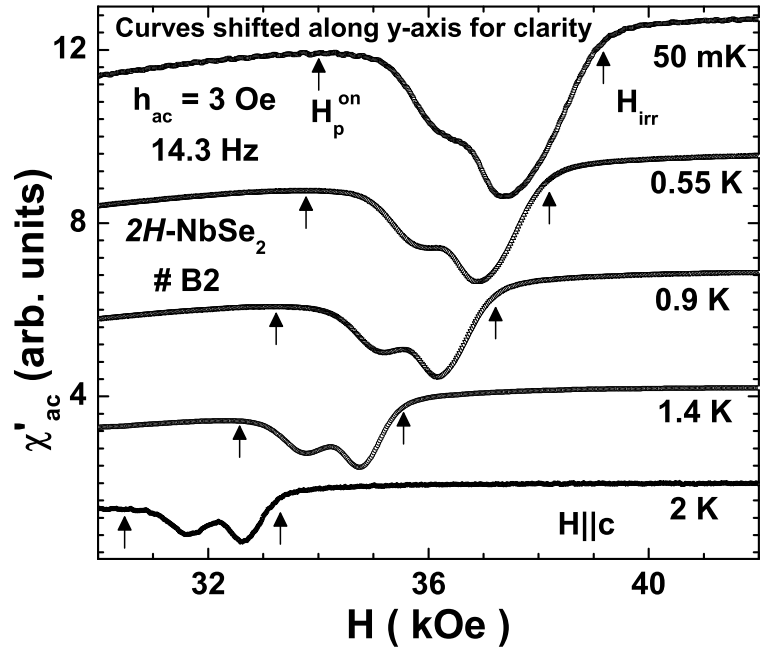}
\end{center}
\caption{Iso-thermal in-phase ac susceptibility ($\chi_{ac}^{ \prime}(H)$) data ($H~\|~c$) in the temperature range 50~mK to 2~K in crystal $B2$ of $2H$-NbSe$_2$.}
\label{fig:4}
\end{figure}

\begin{figure}
\begin{center}
\includegraphics[width=12cm]{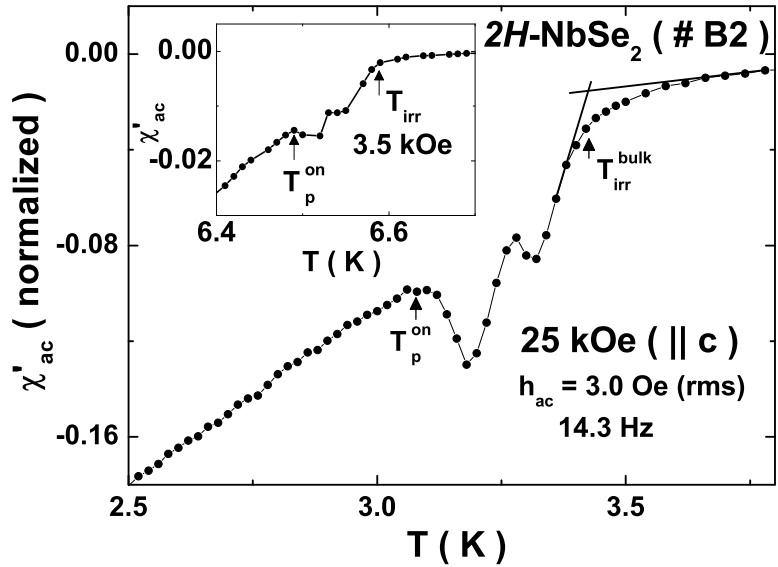}
\end{center}
\caption{In-phase isofield ac susceptibility ($\chi_{ac}^{ \prime}(T)$) data at 25~kOe ($H~\|~c$) in crystal $B2$ of $2H$-NbSe$_2$. An inset panel shows the data at 3.5~kOe ($H~\|~c$).}
\label{fig:5}
\end{figure}

\begin{figure}
\begin{center}
\includegraphics[width=12cm]{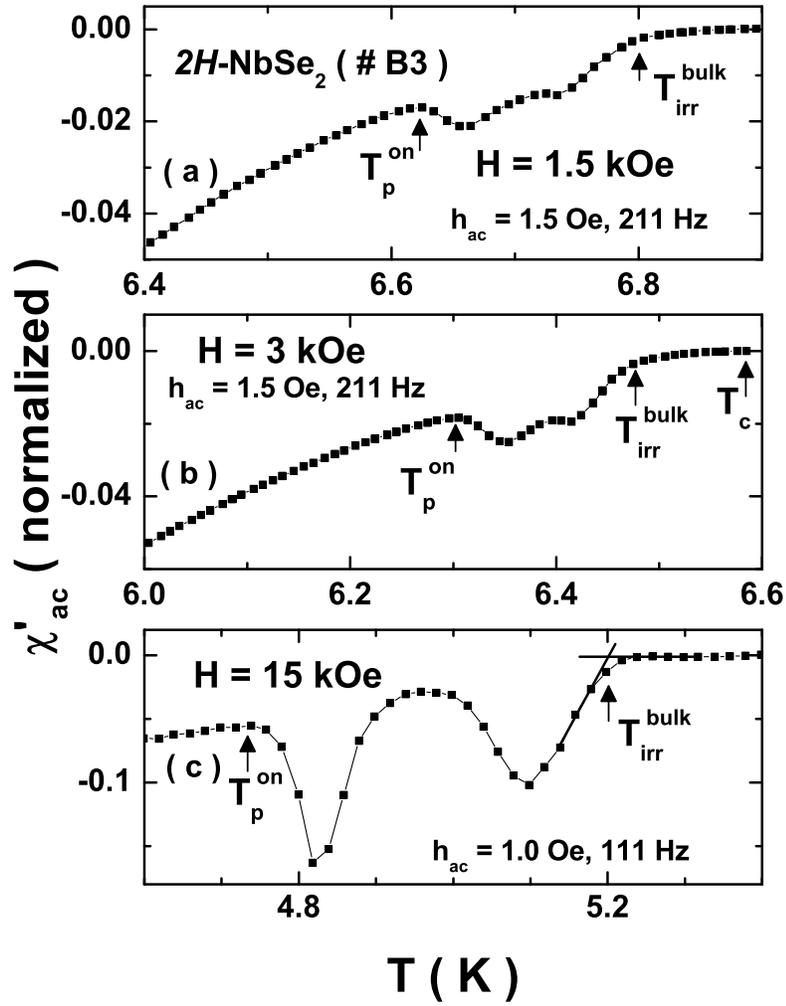}
\end{center}
\caption{In-phase isofield ac susceptibility ($\chi_{ac}^{ \prime}(T)$) data measured at (a)~1.5~kOe, (b)~3~kOe and (c)~12~kOe, respectively for $H~\|~c$ in crystal $B3$ ($T_c(0) \approx 7.2~K$) of $2H$-NbSe$_2$.}
\label{fig:6}
\end{figure}

\begin{figure}
\begin{center}
\includegraphics[width=12cm]{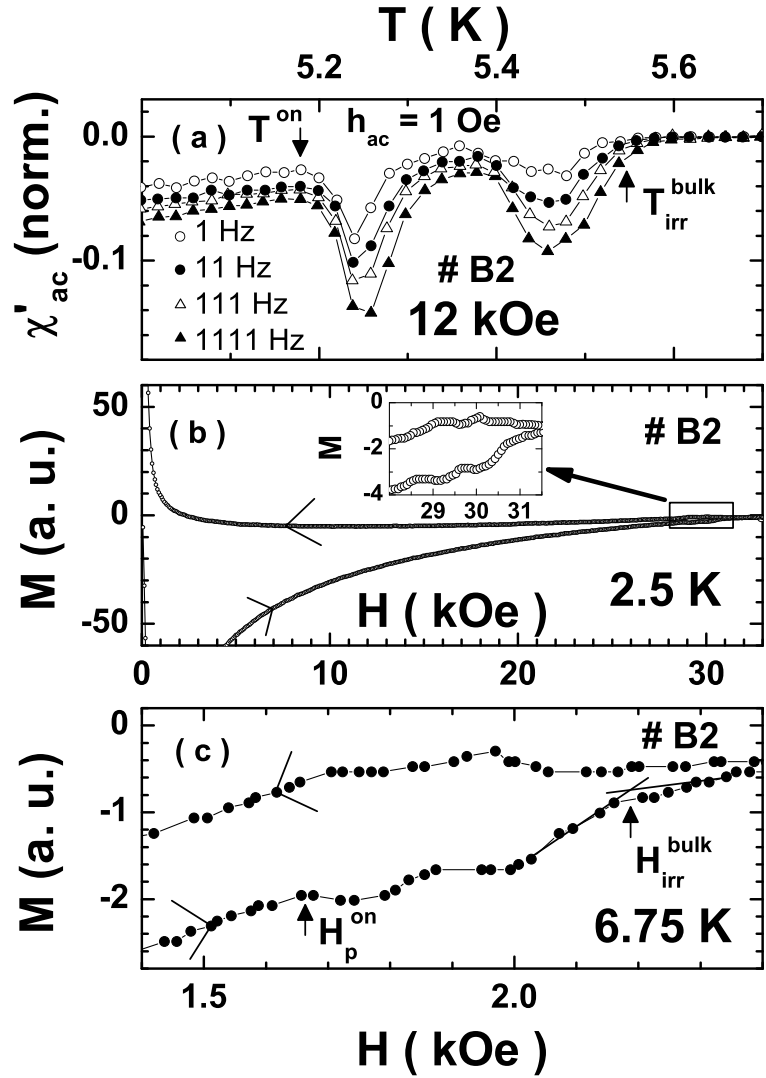}
\end{center}
\caption{Panel (a) shows the in-phase isofield ac susceptibility ($\chi_{ac}^{ \prime}(T)$) data recorded at four frequencies (1~Hz, 11~Hz, 111~Hz and 1111~Hz, respectively) in crystal $B2$ of $2H$-NbSe$_2$ at 12~kOe ($\|~c$). Panels (b) and (c) show portions of the isothermal M-H data in peak effect regime at temperatures 2.5~K and 6.75~K, respectively. An inset in the panel (b) shows an expanded view of the M-H data across peak effect at 2.5~K.}
\label{fig:7}
\end{figure}

\begin{figure}
\begin{center}
\includegraphics[width=12cm]{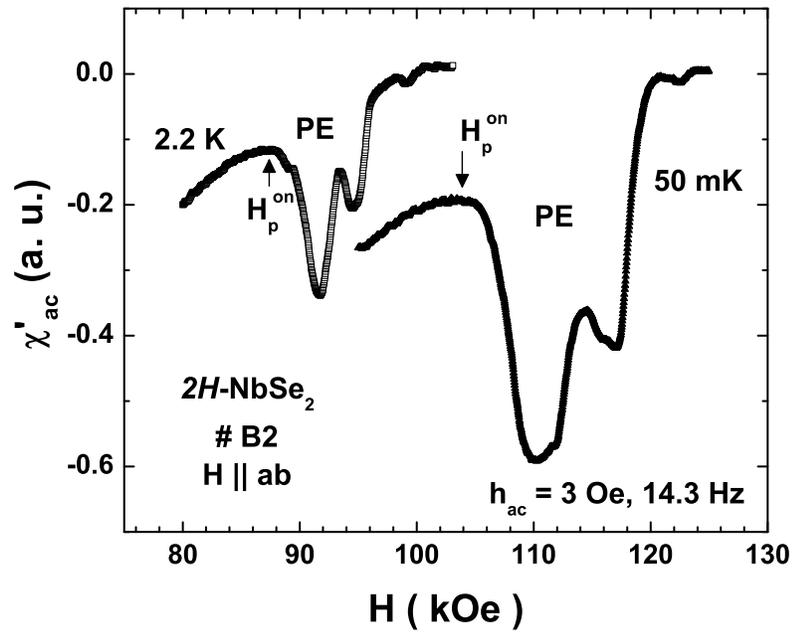}
\end{center}
\caption{In-phase ac susceptibility ($\chi_{ac}^{ \prime}(H)$) data ($H~\|~ab$) at 2.2 K and 50 mK, respectively displaying the two anomalous variations in $J_c(H)$ in crystal $B2$ of $2H$-NbSe$_2$.}
\label{fig:8}
\end{figure}

\begin{figure}
\begin{center}
\includegraphics[width=12cm]{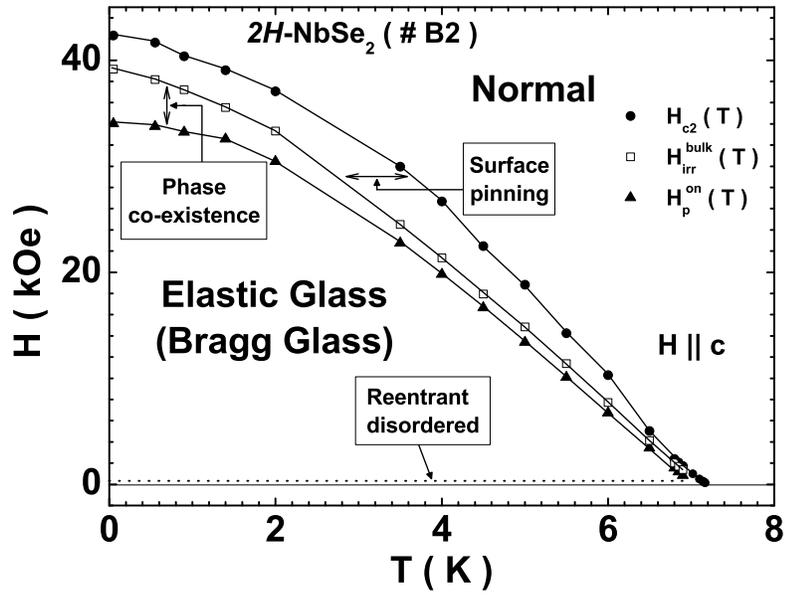}
\end{center}
\caption{A vortex phase diagram constructed from the ac susceptibility data in the crystal $B2$ of $2H$-NbSe$_2$, for $H~\|~c$. The phase space (H, T) regimes corresponding to the reentrant disordered state, elastic glass (i.e., Bragg glass) state and the co-existence of ordered and disordered states have been demarcated. A (H, T) region, where the surface pinning dominates has also been identified.}
\label{fig:9}
\end{figure}

\end{document}